\documentclass[final,3p,times]{elsarticle}
\usepackage{graphicx,color}
\usepackage{amssymb,amsmath,array}
\biboptions{sort,compress}
\journal{Annals of Physics}

\begin{document}

\begin{frontmatter}

\title{On-shell transition of SRG and nuclear systems}

\author[1]{E. Ruiz Arriola}
\author[2]{S. Szpigel}
\author[3]{V. S. Tim\'oteo\footnote{Corresponding author, tel.: +55 11 981 483 747, e-mail address: varese@ft.unicamp.br}}
 
\address[1]{Departamento de F\'isica At\'omica, Molecular y Nuclear and Instituto Carlos I de Fisica Te\'orica y Computacional \\
Universidad de Granada, E-18071 Granada, Spain}

\address[2]{Centro de R\'adio-Astronomia e Astrof\'\i sica Mackenzie, Escola de Engenharia,
Universidade Presbiteriana Mackenzie \\ 01302-907, S\~ao Paulo, SP, Brasil}

\address[3]{Grupo de \'Optica e Modelagem Num\'erica - GOMNI, Faculdade de Tecnologia - FT,
Universidade Estadual de Campinas - UNICAMP \\ 13484-332, Limeira, SP, Brasil}

\begin{abstract}

We make variational estimates for the binding energies of $^2H,~^3H,~^4He,~^{16}O,~^{40}Ca$, 
showing their running with the Similarity Renormalization Group (SRG) cutoff towards the infrared region  
and show the generalized Tjon lines that emerge from the calculations. We infer the SRG evolution 
of the three-body contributions for the $^3H$ and $^4He$ binding energies by computing the two-body 
contributions with a variational approach assuming that four-body forces are negligible. At any given SRG cutoff, 
the three-body contributions may then be inferred as being the experimental value minus the two-body contributions. 
The off-shell / on-shell transition at a critical SRG cutoff $\lambda_c$ drives the behavior of all binding energies and this
scale is the turning point of the generalized Tjon lines. Also, at $\lambda_c$ the ratios between three-body and two-body 
contributions to the binding energies, $B_{\lambda_c}(3) ~/~ B_{\lambda_c}(2)$, are the same in both $^3H$ and $^4He$ 
systems and equal to $1/4$, so that $B_{\lambda_c}(2) ~=~ 4~B_{\lambda_c}(3)$. All calculations are carried out with the 
Idaho-Salamanca N3LO potential, which is evolved with the SRG up to the infrared fixed point ($\lambda \to 0$) in all 
S, P, D, F and G partial-wave channels in order to compute the variational binding energies at several SRG cutoff scales.

\end{abstract}


\end{frontmatter}

\section{Introduction}

Light nuclei are good systems to study few-body forces. Deuteron ($^2H$), Triton ($^3H$) and Helium ($^4He$), 
composed of two, three and four nucleons are then ideal systems to compare different few-body contributions. 
For systems with more nucleons, like oxygen ($^{16}O$) and calcium ($^{40}Ca$), diagonalization of the full hamiltonian 
is a complex task. In the last decade this problem has been overcome by pre-diagonalizing the 
nucleon-nucleon ($NN$) interaction prior to its insertion into the many-body calculations 
(see e.g. the recent work \cite{Maris:2020qne} where this technique is used). 

The pre-diagonalization is achieved by using the so-called Similarity Renormalization Group (SRG), which is a method 
based on a series of continuous unitary transformations that evolve Hamiltonians with a cutoff on energy differences 
preserving the eigenvalues. In particular, SRG transformed two-nucleon interactions also preserve the phase-shifts while giving different contributions to many-body observables. 
The technique was developed independently by Wegner \cite{wegner1994flow} and Glazek \& Wilson \cite{Glazek:1993rc,Glazek:1994qc}
and later applied to nuclear systems in the work of Bogner, Furnstahl and Perry~\cite{Bogner:2006pc}.  

The success of SRG in reducing computational cost for nuclear many-body calculations has motivated a series of studies
in nuclear structure \cite{Bogner:2009bt,Furnstahl:2012fn,Furnstahl:2013oba} that already lasts a decade. 
Typically, two-nucleon forces like phenomenological potentials \cite{Wiringa:1994wb,Stoks:1994wp} or 
effective chiral potentials \cite{Entem:2003ft,Epelbaum:2004fk} are evolved with the SRG up to a point where a minor 
three-nucleon force is enough to provide the correct binding energy. Many works have been carried out in this direction 
showing that indeed great simplification of the many-body problem is achieved by using re-scaled interactions from SRG 
evolution. Moreover, there is a range of SRG cutoffs around $\lambda = 1~{\rm fm}^{-1}$ for which the two-body contribution 
alone is enough to describe the $^3H$ binding energy.

Later, three-nucleon forces were also evolved with the SRG in a series of works \cite{Jurgenson:2009qs,Hebeler:2012pr,Wendt:2013bla} 
showing not only a clean visualization of SRG evolved three-body forces but also the same low-momentum / high-momentum decoupling 
pattern and universality of SRG transformed interactions at small momenta as observed in the case of two-nucleon forces. 

We have also given some contributions on the application of SRG to nuclear physics by studying a leading order effective interaction
renormalized with a subtracted scattering equation \cite{Szpigel:2011bj}, investigating long distance symmetries of the effective 
interactions \cite{Timoteo:2012tt}, comparing implicit and explicit renormalization approaches \cite{Arriola:2013era,Arriola:2014fqa} 
and connecting the infrared limit of the SRG evolution with Levinson's theorem \cite{Arriola:2014aia,Arriola:2016fkr}. Some implications
of on-shell interactions to the nuclear many-body problem, pure neutron matter and the BCS pairing gap 
were presented in Refs. \cite{Arriola:2013gya,Arriola:2014tva,Arriola:2015hra}.

In this work we want to investigate the binding energies of light nuclei and the interplay between contributions from two-body and three-body 
forces as the SRG evolution approaches the on-shell limit. We show that the on-shell transition described in Ref. \cite{Timoteo:2016vlp} drives 
the behavior of the effective interactions as we approach the infrared region of the SRG evolution where the off-shellness of the two-nucleon 
interaction is eliminated by the SRG transformation.   

\section{Two-body and three-body contributions to $^3H$ and $^4He$ binding energies}

We denote by $B_\lambda(N)$ the $N$-body contribution to the binding energy at an SRG cutoff $\lambda$. 
Here we concentrate on the cases $N=2$ and $N=3$ for $^3 H$ ($t$) and the $^4 He$ ($\alpha$).

For a given value of the SRG cutoff $\lambda$ we compute $B_\lambda(2)$ using a variational scheme with harmonic oscillator 
wave functions, minimizing the total energy $E_\lambda = T + V_\lambda$ with respect to the wave function parameter. The interaction $V_\lambda$ is the two-nucleon Idaho-Salamanca N3LO potential \cite{Entem:2003ft}, in a given partial wave, 
evolved to an SRG cutoff $\lambda$.
The number of partial waves we consider depends on the nucleus: only $S$-waves for $^2H$, $^3H$ and $^4He$, up to 
$D$-waves for $^{16}O$ and up to $G$-waves for $^{40}Ca$.

The calculation of the total energy functional for $^2H$, $^3H$ and $^4He$ systems at rest, $\sum_{i=1}^A p_i=0$, 
is simple and can be written as
\begin{eqnarray}
E_\lambda (b)= \min_b \left\{
  (A-1)~ \langle ~1s~ | ~\frac{p^2}{2M}~ | ~1s~ \rangle + \frac{A(A-1)}{2} ~ 
  \langle ~1s~ | ~V^{S}_{\lambda}~ | ~1s~ \rangle_{\rm rel} \right\} \; ,
\label{TEF}
\end{eqnarray}
where $p$ is the nucleon relative momentum (in the CM frame), $M$ is the nucleon mass, $A$ is the nucleus mass number 
and $V^{S}_{\lambda}$ is considered to be the average of the $^1S_0$ and $^3S_1$ waves
\begin{eqnarray}
V^{S}_{\lambda} = \frac12 \left( V^{^1S_0}_{\lambda} + V^{^3 S_1}_{\lambda} \right) \; ,
\end{eqnarray}
and $b$ is the oscillator parameter used to minimize the energy functional $E_\lambda (b)$ from Eq. (\ref{TEF}).
The use of harmonic oscillator wave functions makes the treatment of CM straightforward by subtracting the standard 
term $3\hbar\omega / 2 $ in the expression for the energy, see e.g. Eq. (\ref{TEF}) for A = 2, 3, 4, where the kinetic 
energy term appears with the pre-factor $(A-1)$. A similar $3\hbar\omega/2$ CM subtraction will be implemented for larger 
values of A.

The single particle kinetic energy is given by
\begin{eqnarray}
\left\langle \frac{p^2}{2M} \right\rangle_{1s} = \frac2{\pi}
  \int_0^\infty p^2 \, dp \, \left[\,\varphi_{1s}(p)\,\right]^2 \,
  \frac{p^2}{2M} \; ,
\end{eqnarray}
\noindent
and the potential matrix-elements are defined as
\begin{eqnarray}
\langle \phi_{\rm rel} | V_{\lambda} | \phi_{\rm rel} \rangle =
 \frac{1}{M} \frac{4}{\pi^2} \int_0^\infty  \int_0^\infty dp \, dp' \, p^2 ~ p'^2~  
 \phi_{\rm rel} (p) ~\phi_{\rm rel} (p')  ~ V_{\lambda}(p,p') \;,
\end{eqnarray}
\noindent
where relative wave function is given by
\begin{equation}
\phi_{\rm rel}(p)= \varphi_{\rm 1s} \left(p,b/\sqrt{2}\right)\; .
\end{equation}
\noindent
Eq.~(\ref{TEF}) can be interpreted as the number of nucleon pairs in the $^1S_0$ and $^3S_1$ states, 
being $n^{^1S_0}_t = n^{^3S_1}_t = 3/2$ for the $^3H$ and $n^{^1S_0}_\alpha = n^{^3S_1}_\alpha = 6/2$ 
for the $^4He$.

In the cases of $^{16}O$ and $^{40}Ca$ the matrix elements of the interaction carry contributions from 
higher partial waves and can be written respectively as
\begin{eqnarray}
\langle \phi_{\rm rel} | V_{\lambda} | \phi_{\rm rel} \rangle &=& \frac{1}{M} \frac{4}{\pi^2} \int_0^\infty  \int_0^\infty dp \, dp' \, p^2 ~ p'^2   \\
&\times& 
\left[~\varphi_{1s} (p) ~\varphi_{1s} (p') ~ V^S_{\lambda}(p,p') + \varphi_{2s} (p) ~\varphi_{2s} (p') ~ V^S_{\lambda}(p,p') \right. \nonumber\\
&+& 
\left. \varphi_{1p} (p) ~\varphi_{1p} (p') ~ V^P_{\lambda}(p,p') + \varphi_{1d} (p) ~\varphi_{1d} (p') ~ V^D_{\lambda}(p,p')~\right] \nonumber \; ,
\end{eqnarray}
and 
\begin{eqnarray}
\langle \phi_{\rm rel} | V_{\lambda} | \phi_{\rm rel} \rangle &=& \frac{1}{M} \frac{4}{\pi^2} \int_0^\infty  \int_0^\infty dp \, dp' \, p^2 ~ p'^2  \\
&\times& \left[~\varphi_{1s} (p) ~\varphi_{1s} (p') ~ V^S_{\lambda}(p,p') + \varphi_{2s} (p) ~\varphi_{2s} (p') ~ V^S_{\lambda}(p,p') 
+ \varphi_{3s} (p) ~\varphi_{3s} (p') ~ V^S_{\lambda}(p,p') \right. \nonumber \\
&+& \left. \varphi_{1p} (p) ~\varphi_{1p} (p') ~ V^P_{\lambda}(p,p') + \varphi_{2p} (p) ~\varphi_{2p} (p') ~ V^P_{\lambda}(p,p') 
+ \varphi_{1d} (p) ~\varphi_{1d} (p') ~ V^D_{\lambda}(p,p') \right. \nonumber \\
&+& \left. \varphi_{2d} (p) ~\varphi_{2d} (p') ~ V^D_{\lambda}(p,p') + \varphi_{1f} (p) ~\varphi_{1f} (p') ~ V^F_{\lambda}(p,p')
+  \varphi_{1g} (p) ~\varphi_{1g} (p') ~ V^G_{\lambda}(p,p')  \right] \; , \nonumber
\end{eqnarray}
where $V_\lambda^S,~V_\lambda^P,~V_\lambda^D,~V_\lambda^F,~V_\lambda^G$ are linear combinations of 
($S,P,D,F,G$)-waves which are given explicitly in Ref. \cite{Timoteo:2012tt}.

Within this shell-model calculation scheme the two-body contribution to the variational binding energy is given by
\begin{equation}
B_\lambda(2) = \min_b ~\left\{ \langle  ~ \frac{p^2}{M}~ \rangle + \langle~V_{\lambda}~  \rangle_{{\rm rel}} \right\} \; ,
\label{Bvar}
\end{equation}
which, unlike the experimental value for the two-body binding energy, depends on the SRG cutoff. The three-body contribution 
to the $^3H$ binding energy may be inferred directly since the physical value $B_{\rm exp}$ has to come from two-body plus 
three-body contributions:
\begin{equation}
B^t_\lambda(3) = B^t_{\rm exp} - B^t_\lambda(2)\; .
\label{bt3}
\end{equation}
For the $^4He$ we also have a four-body contribution so that the 3-body contribution reads
\begin{equation}
B^\alpha_\lambda(3) = B^\alpha_{\rm exp} - B^\alpha_\lambda(2) - B^\alpha_\lambda(4) \; .
\label{ba3}
\end{equation}
In this case, the three-body contribution can be deduced only with some assumption on the four-body contribution $B^\alpha_\lambda(4)$. 
When two-nucleon (NN) plus three-nucleon (3N) forces are considered, there are sizeable induced four-body contributions as a side effect
of the SRG flow \cite{Roth2014}. Here we are considering explicitly only two-body forces and the $^4He$ binding energy can be 
reasonably described with just NN+3N forces \cite{Roth2010}, then we neglect the contributions from four-nucleon (4N) forces to keep 
our approach simple. Hence, we assume $B^\alpha_\lambda(4)=0$ for the sake of simplicity.

\section{SRG evolution and the on-shell limit}

The operator form of the SRG equation is given by
\begin{equation}
\frac{d~H_s}{ds} = \left[ \left[ G_s , H_s \right] , H_s \right] \; ,
\label{op}
\end{equation}
with the initial condition $H_0 = H_{s=0}$. Fixed points of Eq. (\ref{op}) are given by stationary solutions ($[[G_s, H_s], H_s] = 0$) 
requiring $[G_s, H_s] = f(H_s)$. Here we assume the usual separation $H_0 = T + V$. For generators which have the
property $\frac{d}{ds} ({\rm tr}~G^2_s) = 0$, and using cyclic properties of the trace and the invariance of ${\rm tr}~(H_s)^n$ one gets
\begin{equation}
\frac{d}{ds} {\rm tr} (H_s - G_s)^2 = -2 ~ {\rm tr}~(i[G_s,H_s])^\dagger (i[G_s,H_s]) \leq 0\; .
\label{fp}
\end{equation}
Since ${\rm tr} (H_s - G_s)^2$ is positive and its derivative is negative, the limit $s \to \infty$ ($\lambda \to 0$) 
exists and correspond to the on-shell limit and any starting interaction $H_0$ is indeed 
diagonalized by the SRG equations with diagonal generators. Thus, the SRG flow can be viewed as a continuous 
diagonalization of $H_0$. The on-shell transition occurs before the on-shell limit as it will be shown in our numerical analysis.

The so-called Wilson generator ($G_s = T$) has a non-negative and decreasing trace of $V^2_s$ which leads 
to an interesting property: the SRG flow with the Wilson generator drives the interaction to smallest possible 
trace of $V^2_s$ having the same spectrum. Isospectrality is clear from the trace invariance 
${\rm tr}~(H_s)^n={\rm tr}~(H_0)^n$.

Defining the Frobenius norm $\phi_\lambda$ for a given SRG cutoff $\lambda$ as \cite{Arriola:2016fkr}
\begin{eqnarray}
\phi_\lambda^2 &=&  {\rm tr}~{V^2_\lambda}  \; , \nonumber \\
                         &=& \left(\frac{2}{\pi}\right)^2 \int_0^\infty dp~p^2 \int_0^\infty dq~q^2 ~V^2_\lambda (p,q) \; ,
\end{eqnarray}
we can state that the fixed-point of the SRG flow for the Wilson generator is the interaction with the smallest norm.
This means we can use the Frobenius norm as a metric to quantify the off-shellness of a potential or even to determine 
how close (or far) we are from the on-shell limit, where the interaction is completely diagonal. 

The way an initial Hamiltonian with large norm (more off-shellness) flows with the SRG towards the on-shell limit (no off-shellness) 
resembles the same phase transition behavior as observed in the two-flavor NJL model or in the two-dimensional Ising model. 
The phenomenon can be described by defining an order parameter $\beta$ for the off-shell / on-shell transition as the derivative 
of the Frobenius norm of the potential with respect to the SRG cutoff \cite{Timoteo:2016vlp}, 
\begin{equation}
\beta_\lambda = \frac{\partial \phi_\lambda}{\partial \lambda} \; . 
\end{equation}

The pseudo-critical value of the SRG cutoff $\lambda_c$ where the on-shell transition occurs is found by looking at 
\begin{eqnarray}
\eta_\lambda &=& \frac{\partial \beta_\lambda}{\partial \lambda} \; ,  
\end{eqnarray}
which may be regarded as a \emph{similarity susceptibility} of the two-nucleon interaction and present a peak where the 
on-shell transition takes place. 

The family of isospectral interactions $V_\lambda$ obtained from the N3LO chiral potential is obtained by solving
the SRG flow equation numerically. For each partial wave, the flow equation is given by
\begin{eqnarray}
\frac{dV_s(p,p')}{ds}=-(\epsilon_p-\epsilon_p')^2 \; V_s(p,p')+\frac{2}{\pi} \int_{0}^{\infty}dq \; q^2\;
(\epsilon_p+\epsilon_p'-2 \epsilon_q)\; V_s(p,q)\; V_s(q,p') \; ,
\label{flowWil}
\end{eqnarray}
where  $s = 1 / \lambda^4$ is the flow parameter as a function of the SRG cutoff $\lambda$ and the initial condition 
at $s = 0$ ($\lambda = \infty$) is the N3LO potential $V_\chi$ in a given partial wave ${}^{2S+1} L_{~J}$:
\begin{eqnarray}
V_{s=0} (p,p') = V_{\lambda=\infty} (p,p') = V_\chi (p,p') \; .
\label{ic_flow}
\end{eqnarray}

All solutions of the flow equation represent a two-nucleon force $V_\lambda$ that provides the same two-nucleon phase-shifts 
and eigenvalues but gives different contributions for binding energies of light nuclei when inserted in Eq. (\ref{Bvar}). 

\section{Numerical results}

We start by presenting some of the effective interactions used in this work. We considered the Idaho-Salamanca 
N3LO potential \cite{Entem:2003ft} and its evolution with the SRG for cutoffs in the range 
$0.1 \leq\lambda\leq 2.0~{\rm fm}^{-1}$. The calculations with an S-wave toy model \cite{Arriola:2014fqa,Arriola:2016fkr} 
showed that when the interaction has a short tail in momentum space we can perform the evolution up to the on-shell 
limit with small grid configurations. And this is also the case of the chiral N3LO potential: its exponential regulating function 
with $\Lambda = 500 ~{\rm MeV}$ ($\sim 2.5~{\rm fm}^{-1}$) provides good convergence in our calculations of variational 
binding energies and two-nucleon phase-shifts if we consider a maximum momentum $p_{\rm max} = 4 ~{\rm fm}^{-1}$ 
and $N = 30$ grid points. The advantages of the N3LO interaction are that it provides good description in all partial waves
required in our calculations and the S-waves contain both attraction and repulsion unlike the gaussian toy model which contains
only attraction.

While a perfect description of the $^1S_0$ phase-shifts with the N3LO potential requires a larger maximum momentum 
in the integral equation ($p_{\rm max} \sim 20 ~{\rm fm}^{-1}$), a large $p_{\rm max}$ implies in more grid points which, 
in turn, makes the SRG evolution towards the infrared limit extremely expansive computationally. While there exists the 
possibility of advantageously reducing the number of grid points \cite{Gomez-Rocha:2019xum, RTS2021}, here we are 
interested in qualitative results from variational calculations, we restrict the momentum grid with 
$p_{\rm max} = 4 ~{\rm fm}^{-1}$ which provides good results with $N = 20$ and $N = 30$ points. 

\begin{figure*}[t]
\begin{center}
\includegraphics[scale=0.3]{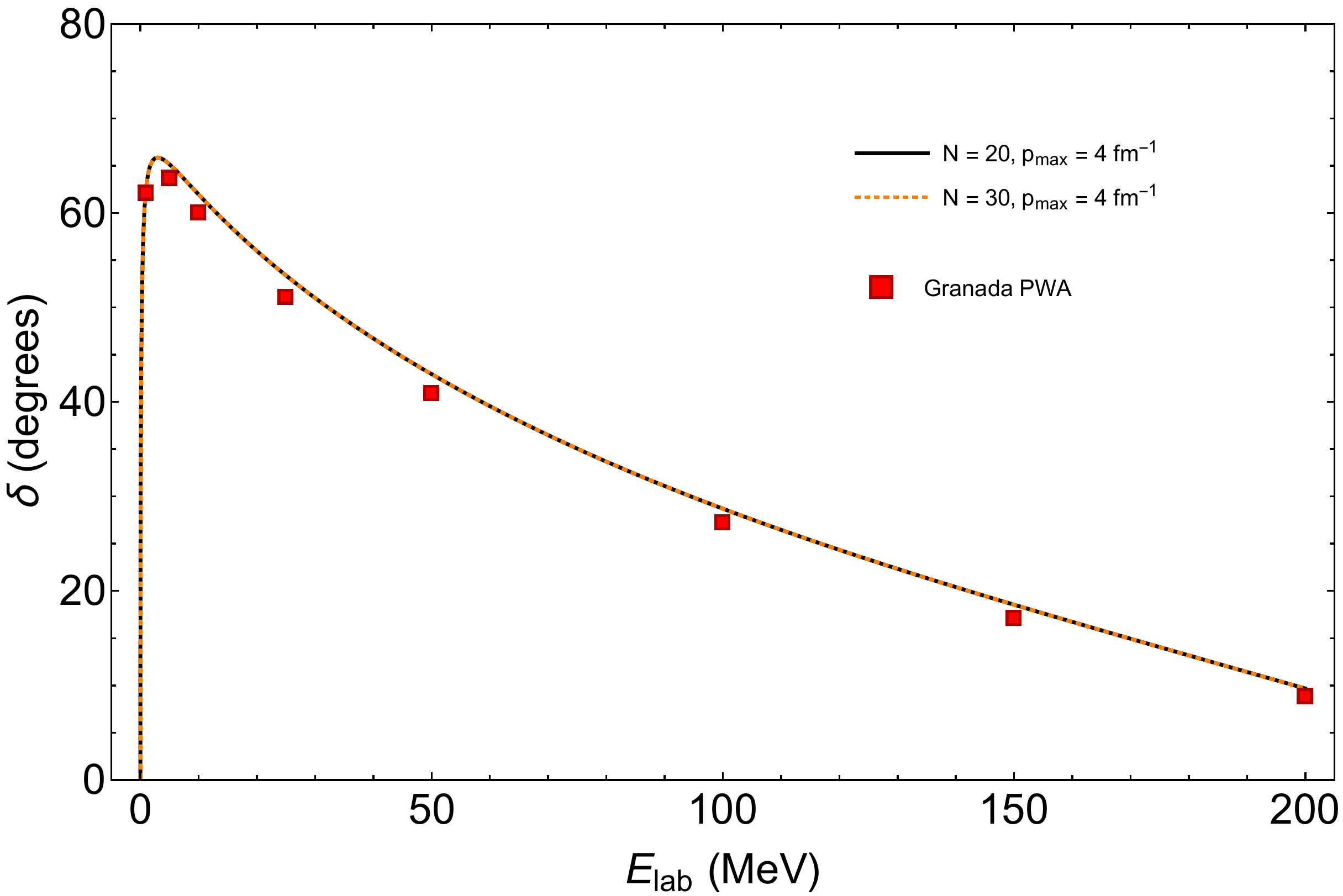}  
\end{center}
\caption{Phase-shifts from the N3LO chiral potential in the $^1S_0$ channel compared to the Granada partial wave analysis \cite{ugrpwa}.}
\label{fig1}
\end{figure*}

The $^1S_0$ phase-shifts for small grid configurations are displayed in Fig. \ref{fig1}, where we can see a reasonable 
description of the Granada partial wave analysis \cite{ugrpwa} considering we are interested in variational binding energies 
which are not accurate. Also, $N = 20$ or $N = 30$ grid points give the same set of variational binding energies for different 
SRG cutoffs. We then use $N=20$ points in order to reduce the runtime to evolve the potential with the the SRG up to the infrared 
limit for all channels up to G-waves, here set to $\lambda_0 = 0.1~{\rm fm}^{-1}$ and numerically shown to be a fixed point \cite{Arriola:2016fkr}.  

\begin{figure*}[b]
\begin{center}
\includegraphics[scale=0.45]{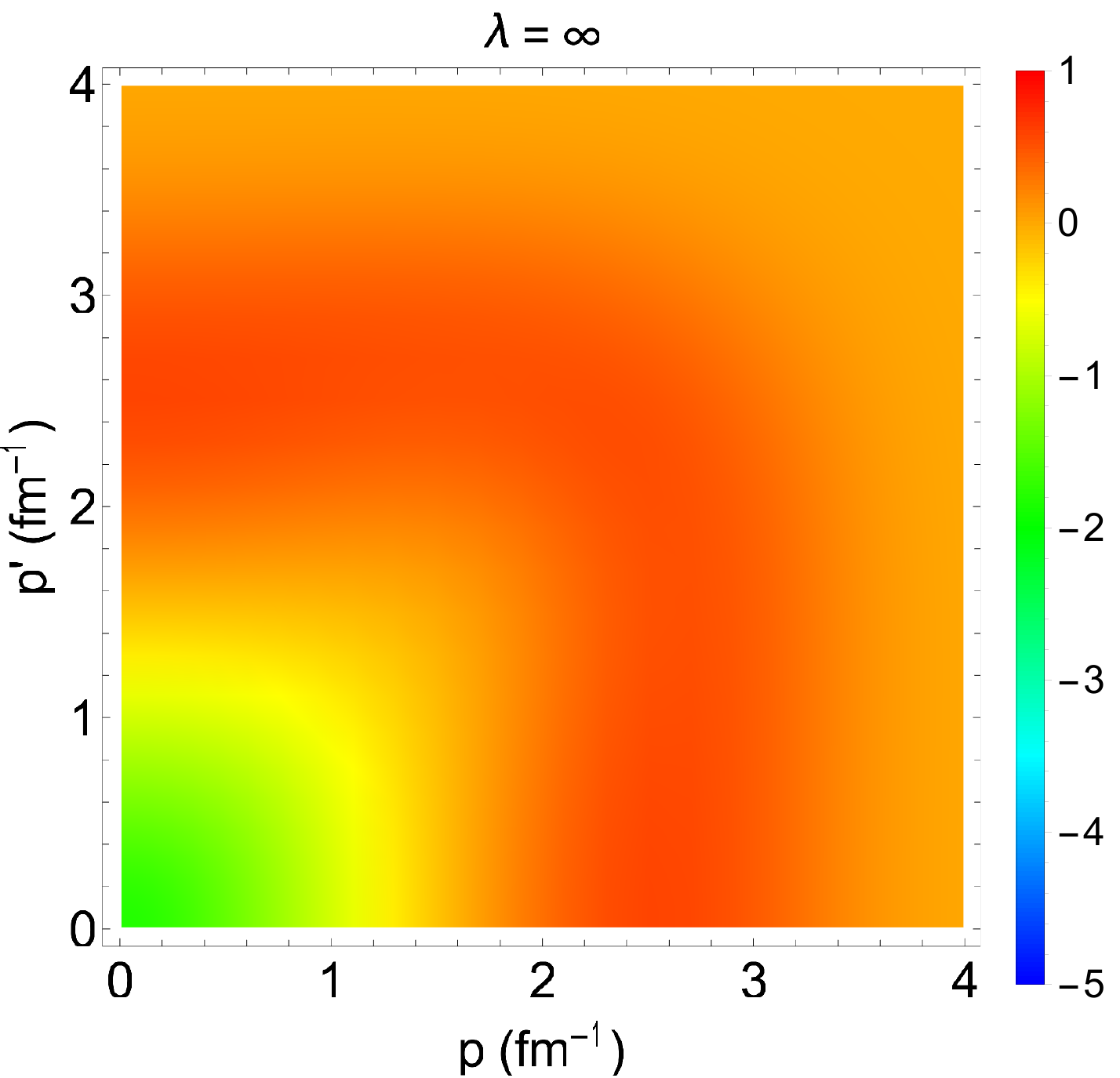} \hspace*{1cm}
\includegraphics[scale=0.45]{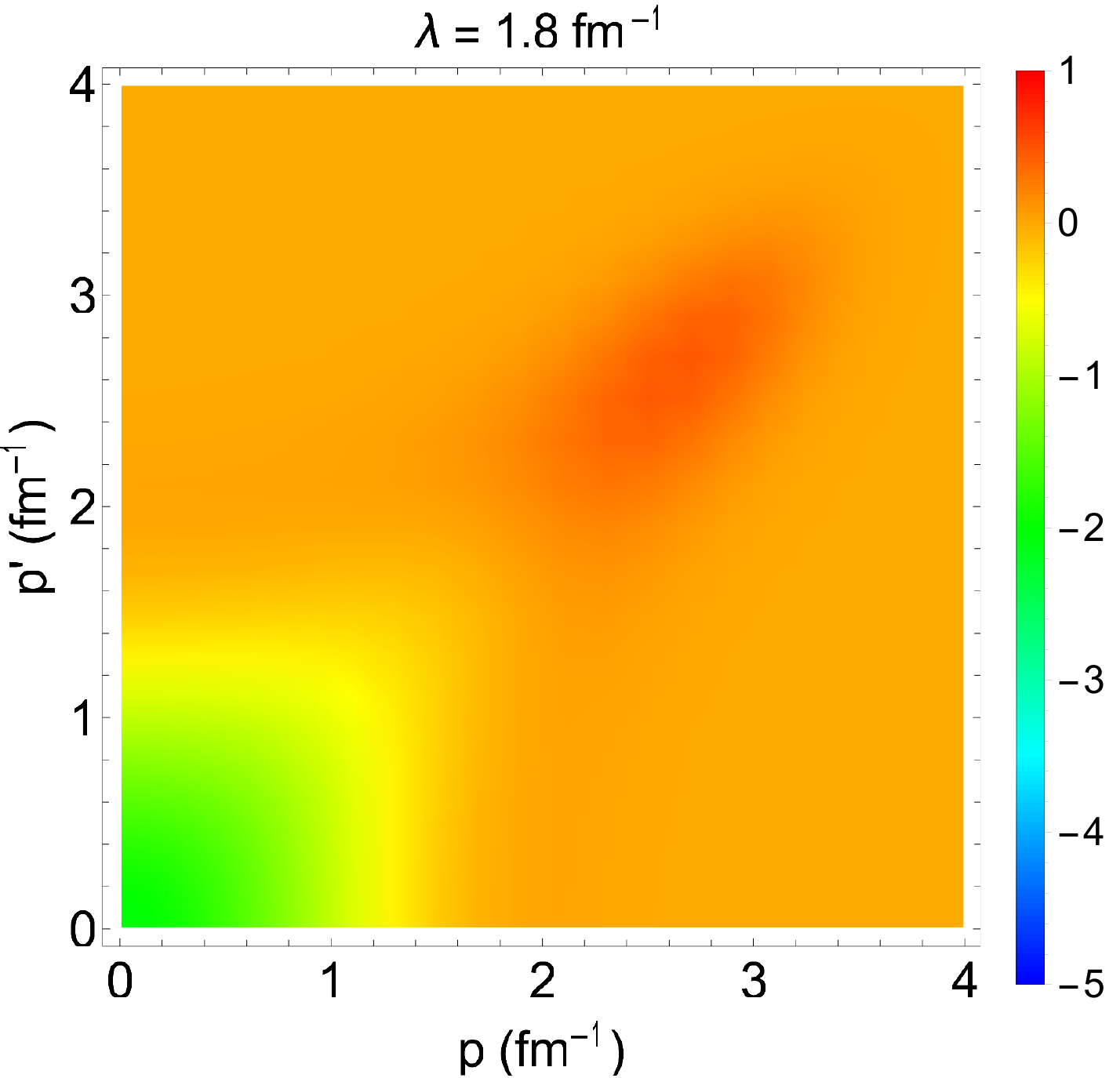} \\ \vspace*{1cm}
\includegraphics[scale=0.45]{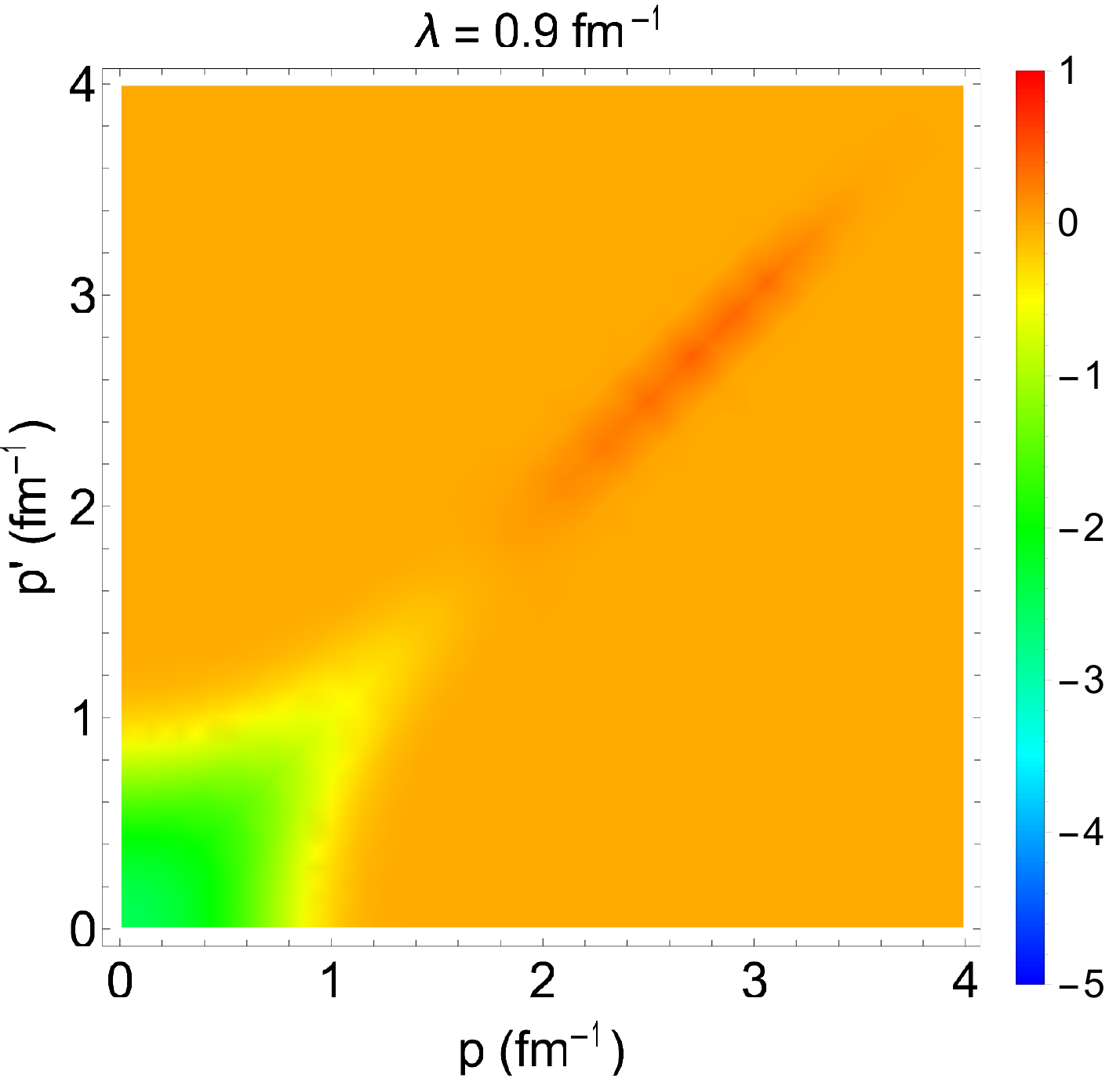} \hspace*{1cm}
\includegraphics[scale=0.45]{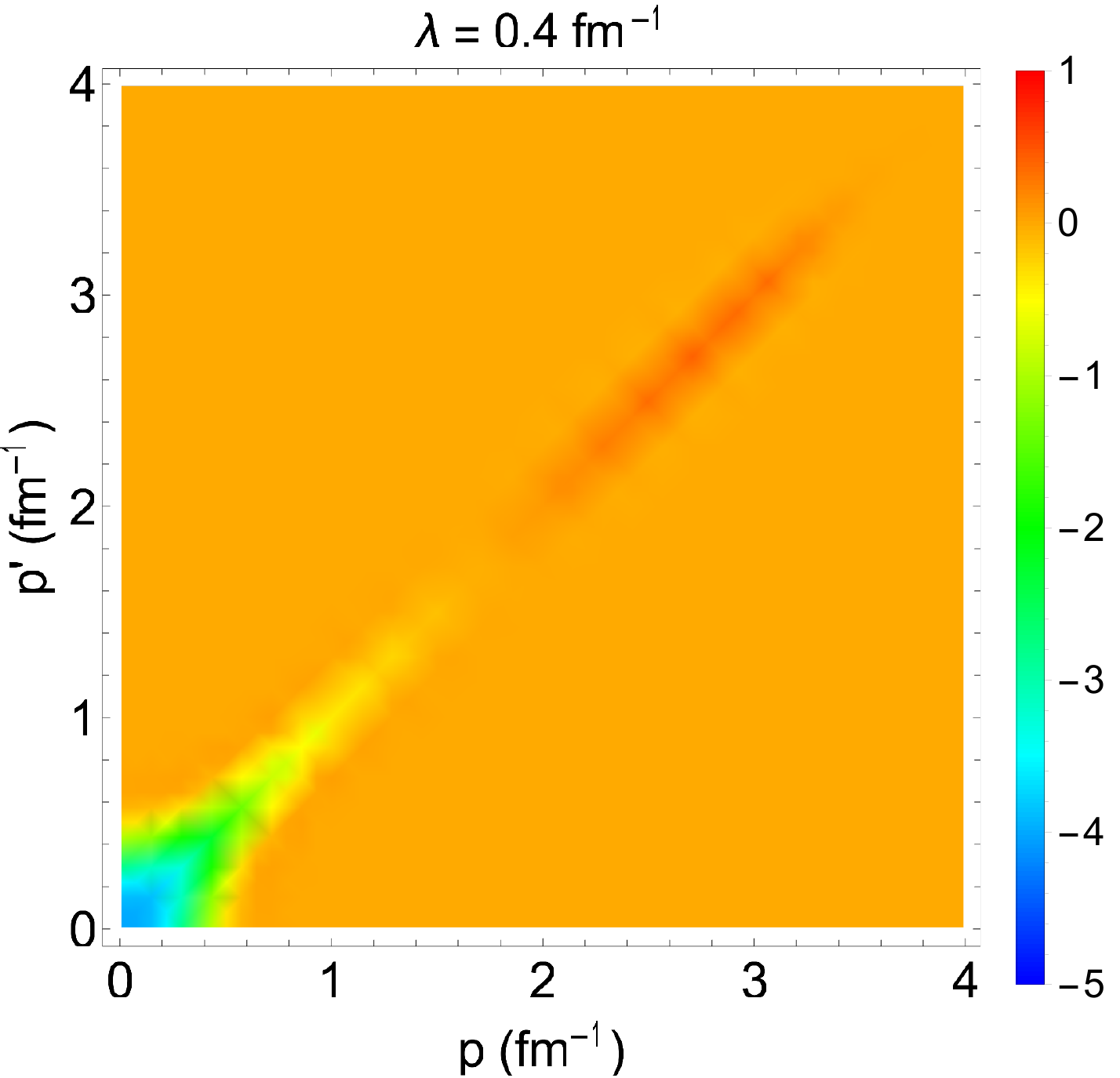}
\end{center}
\caption{Evolution with the SRG of the chiral N3LO potential, given in fm, in the $^1S_0$ channel for selected values of the SRG cutoff.
Similar flows towards a diagonal form occur in all partial waves and here we consider up to G-waves.}
\label{fig2}
\end{figure*}

\begin{figure*}[t]
\begin{center}
\includegraphics[scale=0.45]{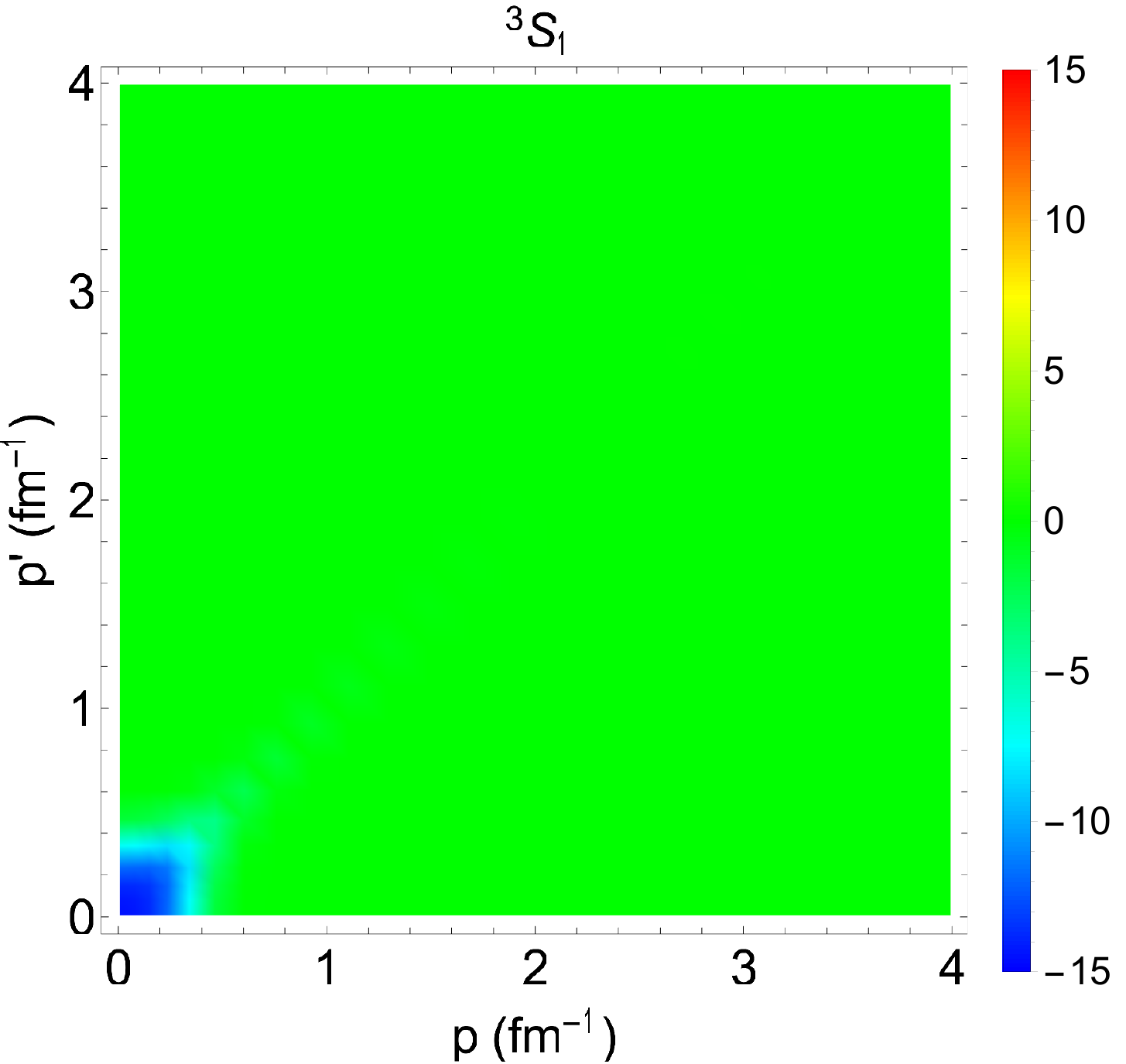} \hspace*{1cm}
\includegraphics[scale=0.45]{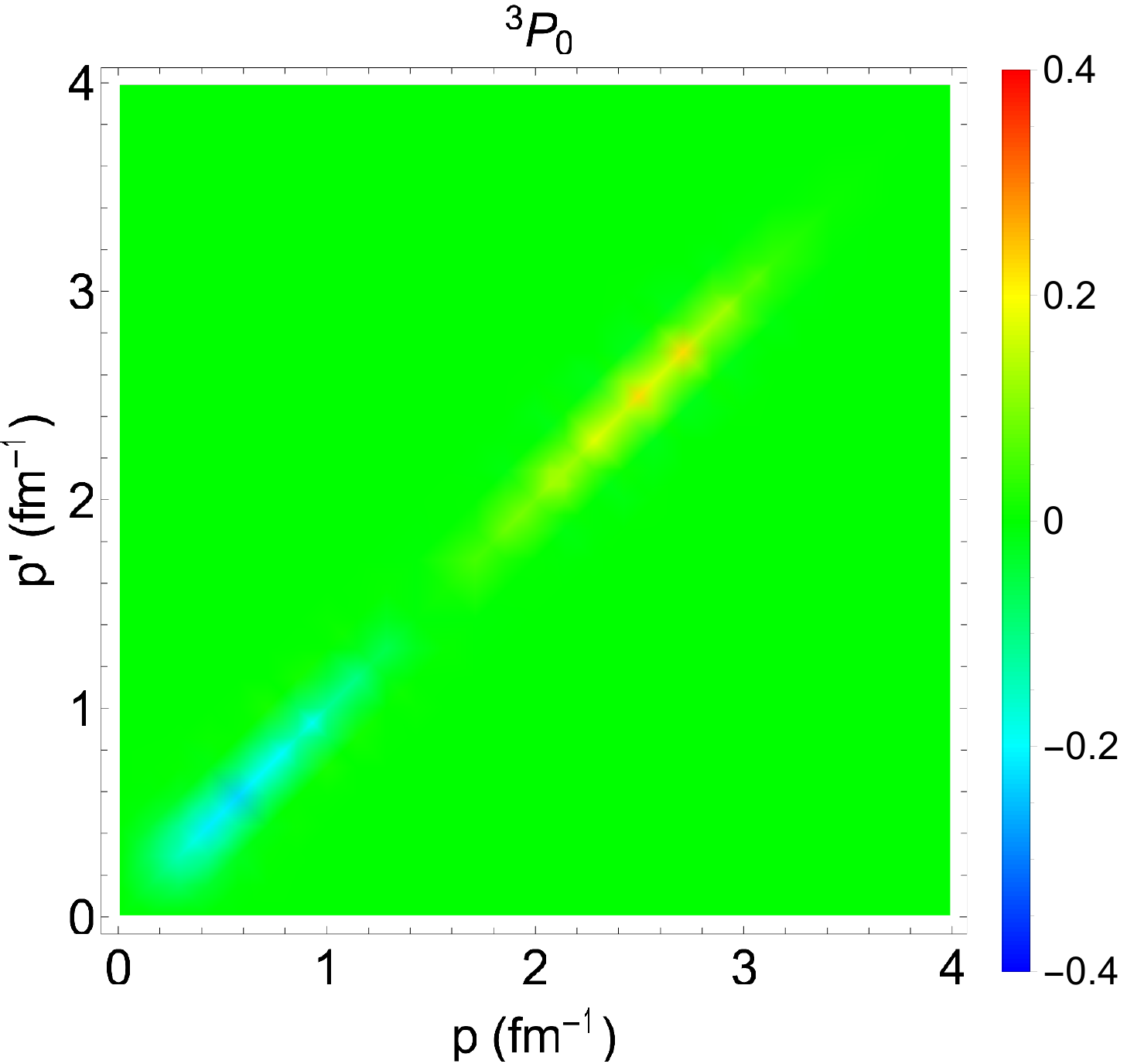} \\ \vspace*{1cm}
\includegraphics[scale=0.45]{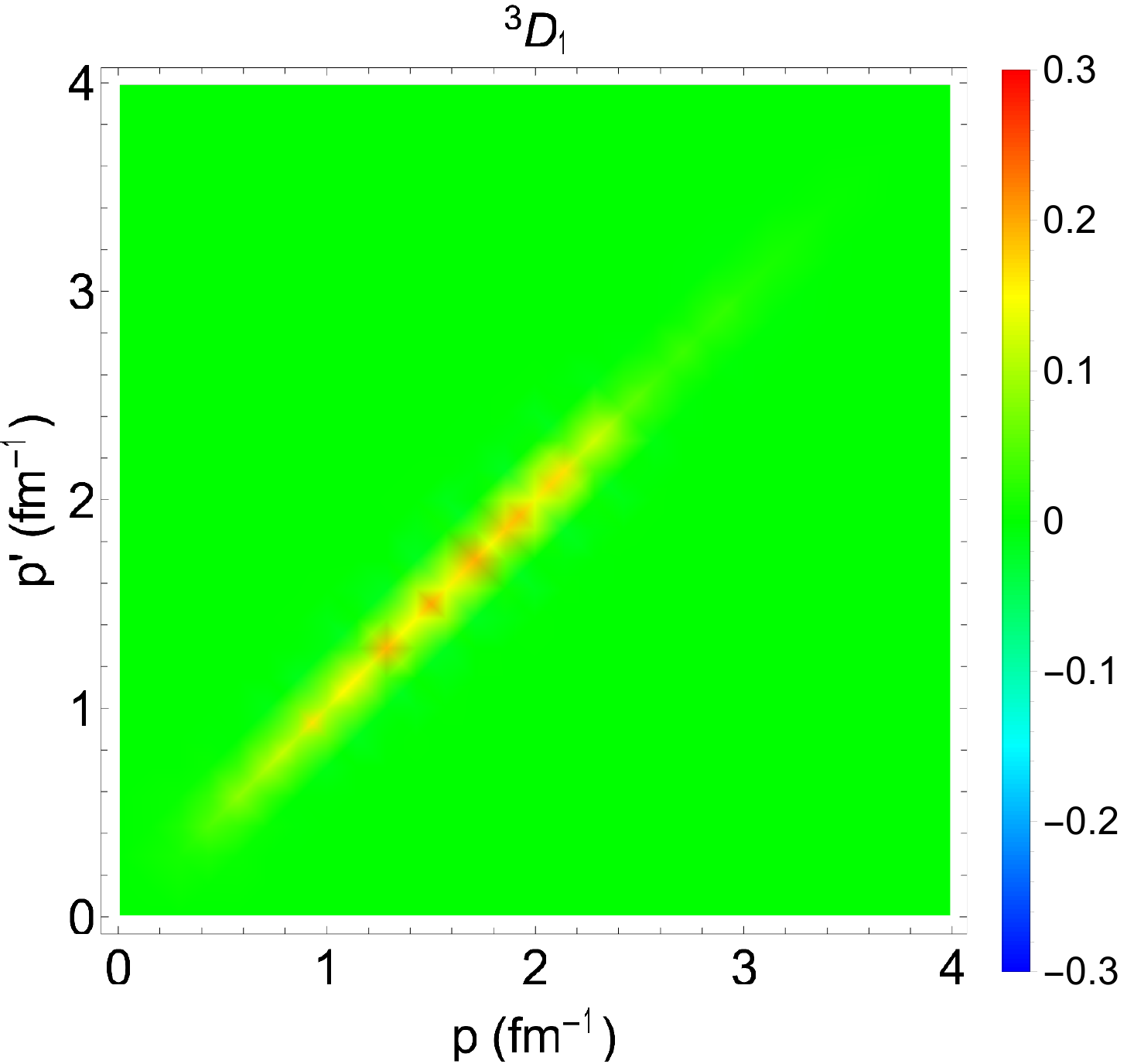} \hspace*{1cm}
\includegraphics[scale=0.45]{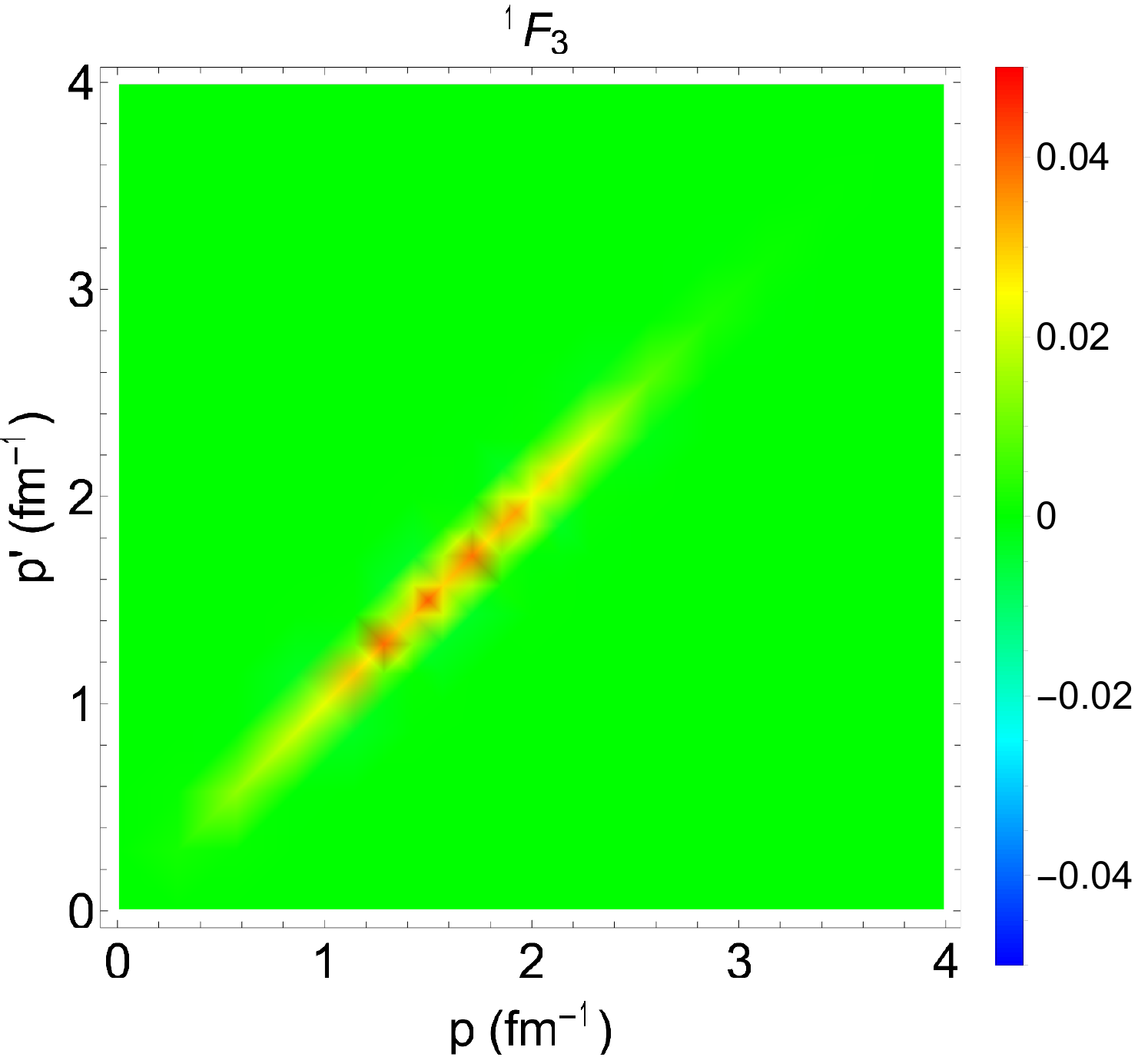}
\end{center}
\caption{Evolution with the SRG of the chiral N3LO chiral, given in fm, in selected channels for $\lambda = 0.4~{\rm fm}^{-1}$.}
\label{fig3}
\end{figure*}

In Fig. \ref{fig2} we display the two-nucleon effective interaction for selected SRG cutoffs in the $^1S_0$ channel. The case 
$\lambda=\infty$ corresponds to the original N3LO potential without any SRG evolution. The other values of $\lambda$ 
correspond to cutoffs before, at and after the on-shell transition. In Fig. \ref{fig3} we show selected channels of the
N3LO potential at $\lambda =  0.4~{\rm fm}^{-1}$, where it is evident that the effective interactions in higher partial waves 
are practically diagonal.

\begin{figure*}[b]
\begin{center}
\includegraphics[scale=0.3]{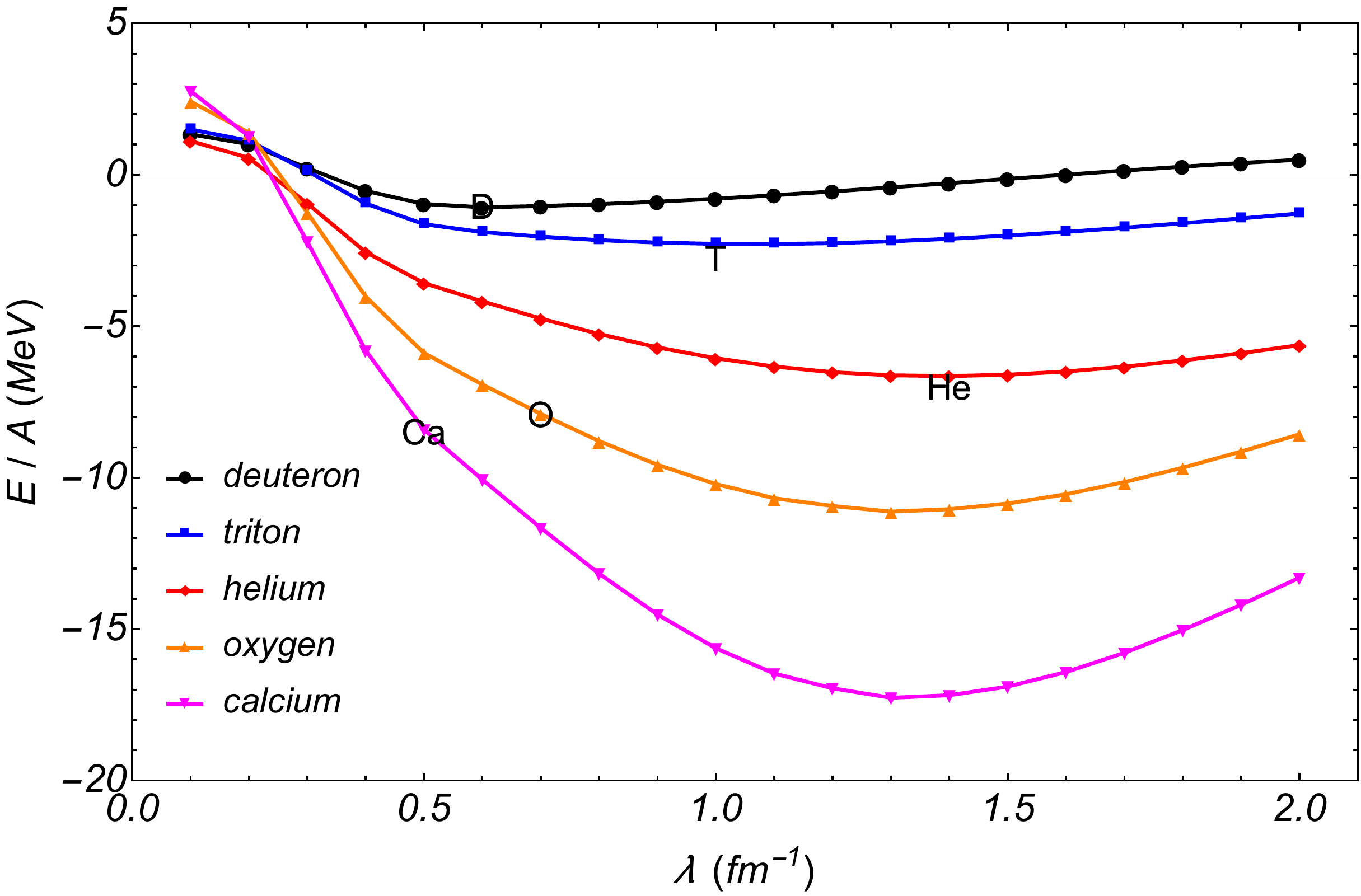} \hspace*{1cm}
\includegraphics[scale=0.3]{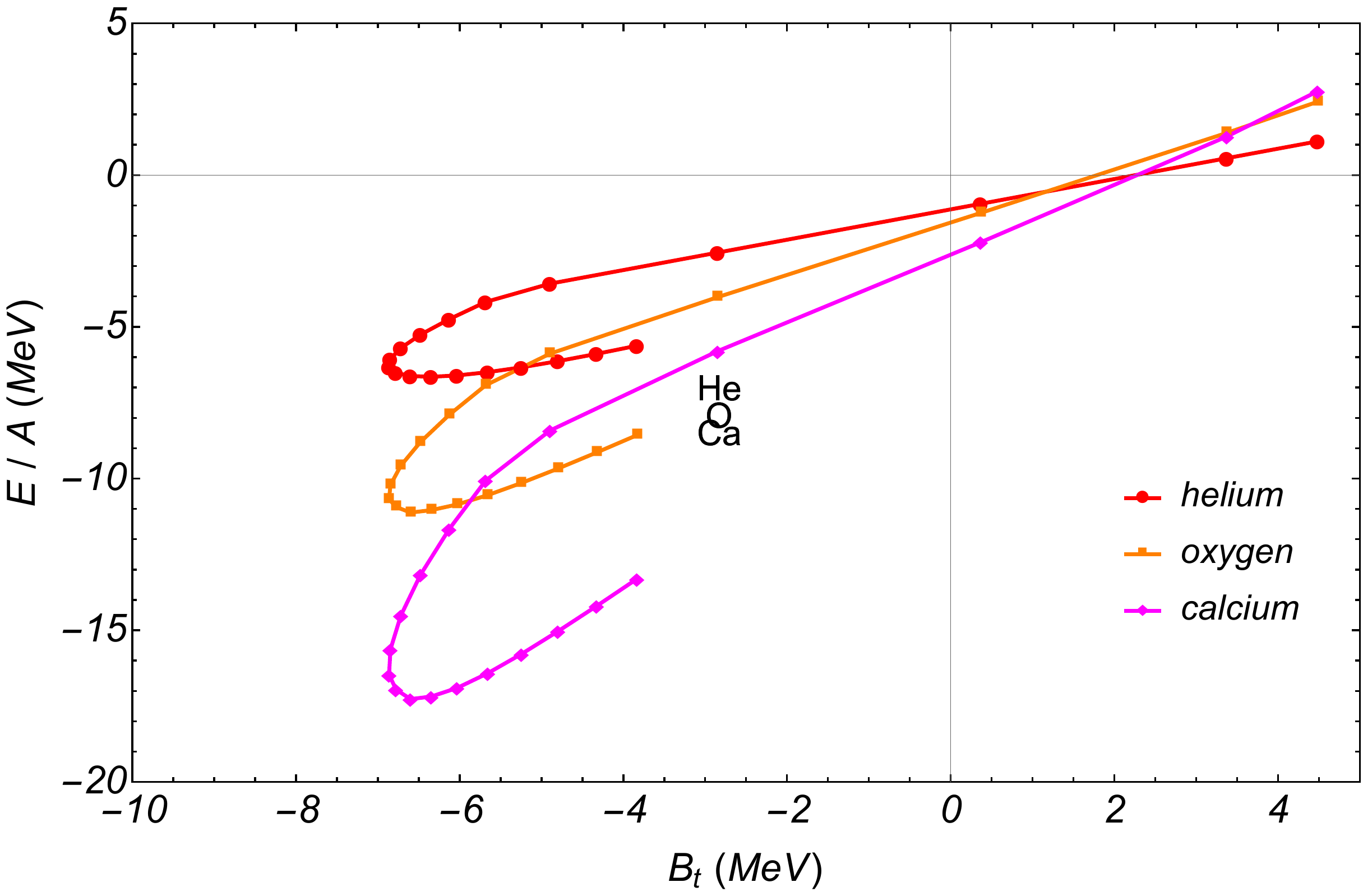}
\end{center}
\caption{Running of the two-body variational binding energies per nucleon with the SRG cutoff including the infrared region (left) 
and the generalized Tjon lines for $^4He$, $^{16}O$ and $^{40}Ca$ where each point in the curves corresponds to a different
$\lambda$ (right). The experimental values are marked with the labels Ca, O, He, T and D.}
\label{fig4}
\end{figure*}

The first analysis we make for few light nuclei is the dependence of their binding energy per nucleon on the SRG
cutoff when only two-body forces are taken into account. This is shown in the left panel of Fig. \ref{fig4} where we can 
clearly observe the running of all binding energies with the SRG evolution and that they all come closer to their respective 
experimental value for some choice of a phenomenological SRG cutoff $\lambda_{\rm ph}$, which is different for each nuclei 
($\lambda_{\rm ph} = 0.6~{\rm fm}^{-1}$ for $^2H$, $\lambda_{\rm ph} = 1.0~{\rm fm}^{-1}$ for $^3H$, 
$\lambda_{\rm ph} = 1.4~{\rm fm}^{-1}$ for $^4He$, $\lambda_{\rm ph} = 0.7~{\rm fm}^{-1}$ for $^{16}O$ and 
$\lambda_{\rm ph} = 0.5~{\rm fm}^{-1}$ for $^{40}Ca$). 

From the binding energies we can construct generalized Tjon lines by plotting the binding energy
per nucleon $E/A$ against the $^3H$ binding energy $B_t$. This is displayed in the right panel of Fig. \ref{fig4} where
we observe that the experimental values are off the numerical results since the variational calculation of the binding energies 
is not accurate. Note that results depicted in Fig. \ref{fig4} show only the contribution from SRG evolved two-nucleon interactions 
and does not include many-body forces of any kind.

Now that we have computed the two-body contribution to the binding energies for several values of $\lambda$,
we can infer the three-body contribution at each SRG cutoff by simply applying Eqs. (\ref{bt3}) and (\ref{ba3}). 
In Fig. \ref{fig6} we show the contributions from two-nucleon and three-nucleon forces to the binding
energy of the $^3H$ (left panel) and $^4He$ (right panel) for several values of $\lambda$, with the respective 
experimental value. 

Since the physical value for the binding energy is fixed, the three-body contribution is smaller 
when the two-body contribution is larger, and this happens at $\lambda = 1.1~{\rm fm}^{-1}$ for the $^3H$ and at 
$\lambda = 1.4~{\rm fm}^{-1}$ for the $^4He$, which means that a reasonable description is obtained including only 
two-nucleon forces when $B_n(3) \sim 0$. At $\lambda \sim 0.5~{\rm fm}^{-1}$, where the black and blue curves 
cross each other, the contributions from both two-nucleon and three-nucleon forces are equivalent in both $^3H$ and 
$^4He$ systems. Also, at $\lambda \sim 0.3~{\rm fm}^{-1}$ where the black and red curves cross each other, we have 
the binding energies for $^3H$ and $^4He$ described only by three-nucleon forces since $B_n(2) = 0$. 

Having determined how three-body contributions depend on the SRG cutoff for both $^3H$ and $^4He$, 
we may compare the two-body and three body contributions by looking at the ratio $B(3)/B(2)$. This is shown
in Fig. \ref{fig6} for both $^3H$ and $^4He$. The smaller is $B(3)$ as compared to $B(2)$ the better is the 
description of the binding energy with only SRG evolved two-body forces, so the minimum of $B(3)/B(2)$
for $^3H$ and $^4He$ indicates the values of the SRG cutoff at which the description with only
two-body forces are best for the triton ($\lambda_{\rm 2B} = 1.05~{\rm fm}^{-1}$) and for the helium 
($\lambda_{\rm 2B} = 1.40~{\rm fm}^{-1}$). 

\begin{figure*}[h]
\begin{center}
\includegraphics[scale=0.3]{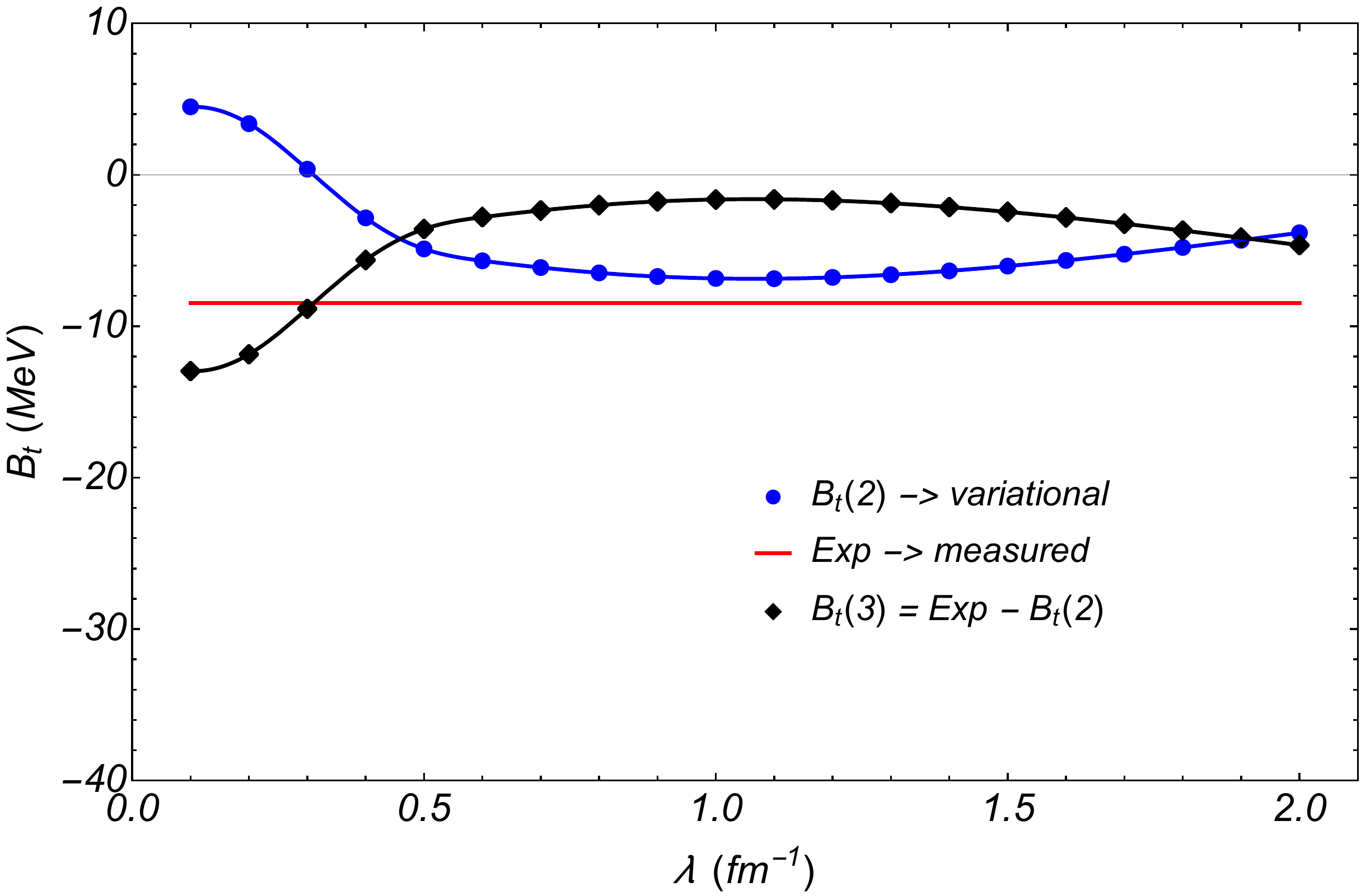} \hspace*{1cm}
\includegraphics[scale=0.3]{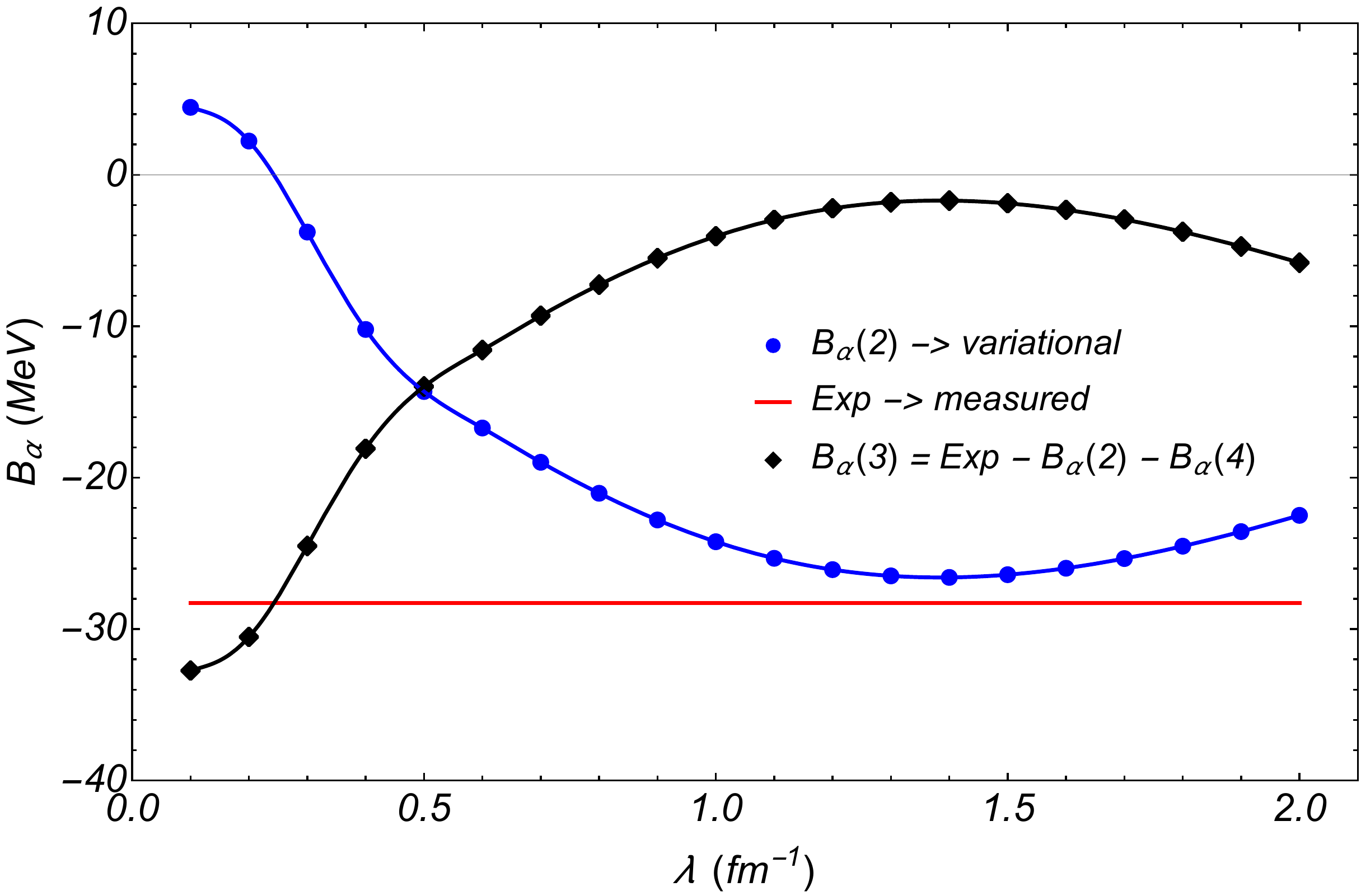}
\end{center}
\caption{Calculated two-body and inferred three-body contributions for the $^3H$ binding energy as functions of the SRG cutoff (left) 
and the same for the $^4He$ binding energy, disregarding contributions from four-body forces by setting $B_\alpha (4) = 0$ (right). 
Red lines represent the experimental values for the binding energies.}
\label{fig5}
\end{figure*}

\begin{figure*}[h]
\begin{center}
\includegraphics[scale=0.4]{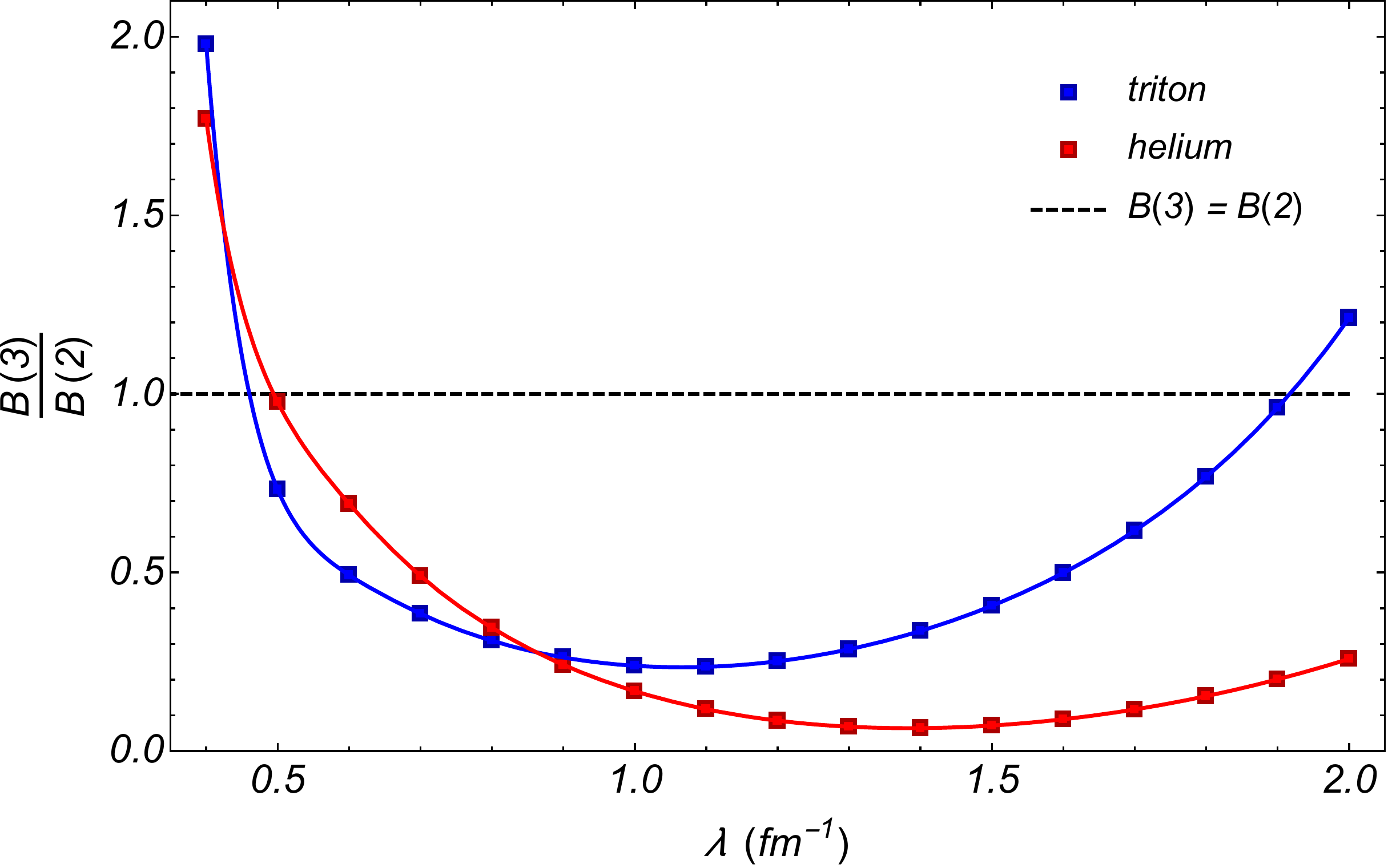}  
\end{center}
\caption{Ratio between the inferred three-body contributions and the calculated two-body contributions for $^3H$ (blue)  
and $^4He$ (red). The black dashed line correspond to equal contributions from two-body and three-body forces, 
$B(3)=B(2)$. For $\lambda < 0.4~{\rm fm}^{-1}$ the triton becomes unbound.}
\label{fig6}
\end{figure*}

Also, Fig. \ref{fig6} shows that $B(3)/B(2)$ is the same in both triton and helium when $\lambda \sim 0.9~{\rm fm}^{-1}$, 
indicating that this value of the SRG cutoff is special since it is also the value where the generalized Tjon lines present a 
sharp turn (see the right panel of Fig. \ref{fig4}). For $\lambda < 0.4~{\rm fm}^{-1}$ the triton start to be unbound since 
$E/A$ becomes positive (see Fig. \ref{fig5}).

\begin{figure*}[h]
\begin{center}
\includegraphics[scale=0.3]{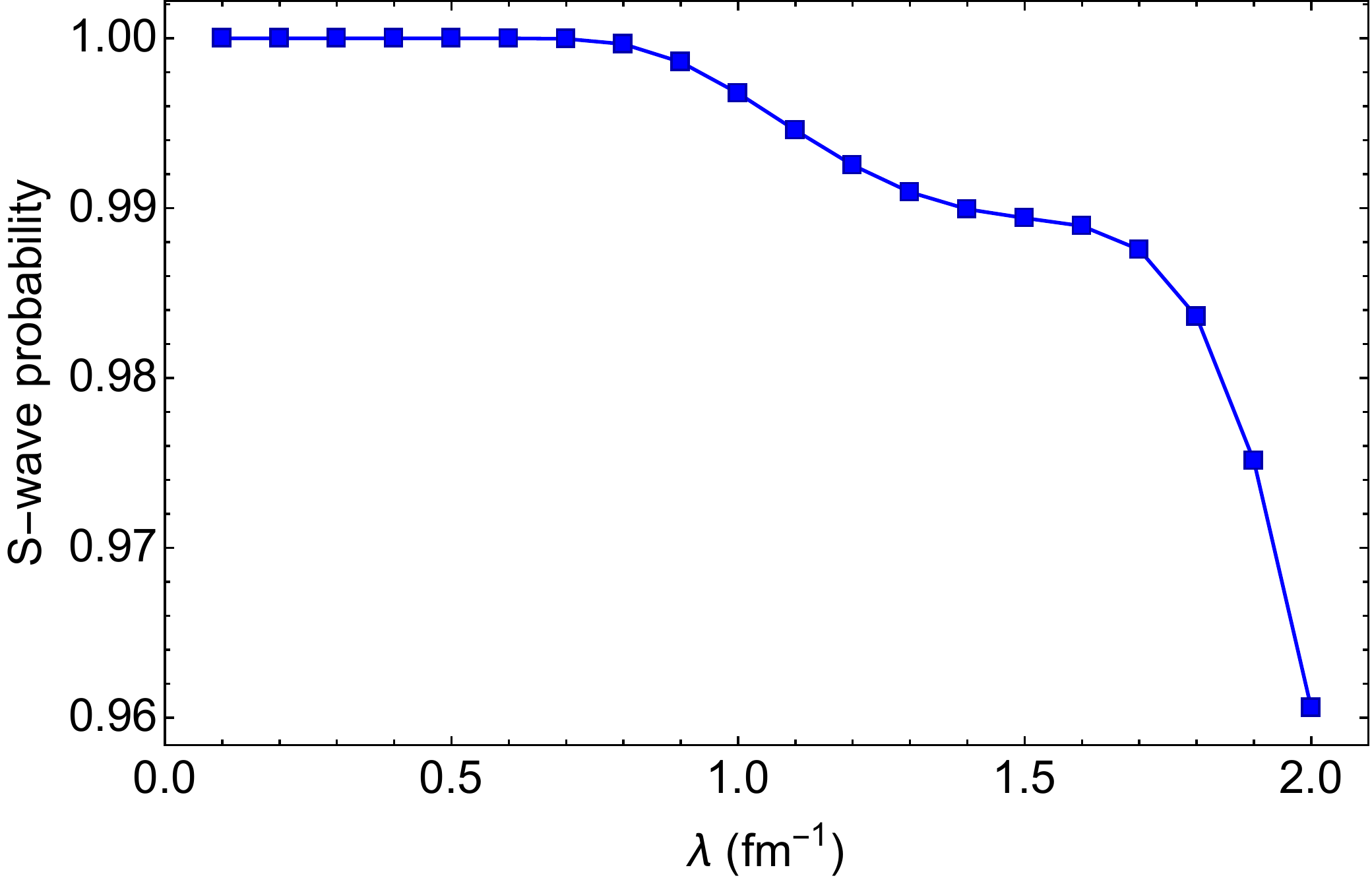} \hspace*{1cm}
\includegraphics[scale=0.3]{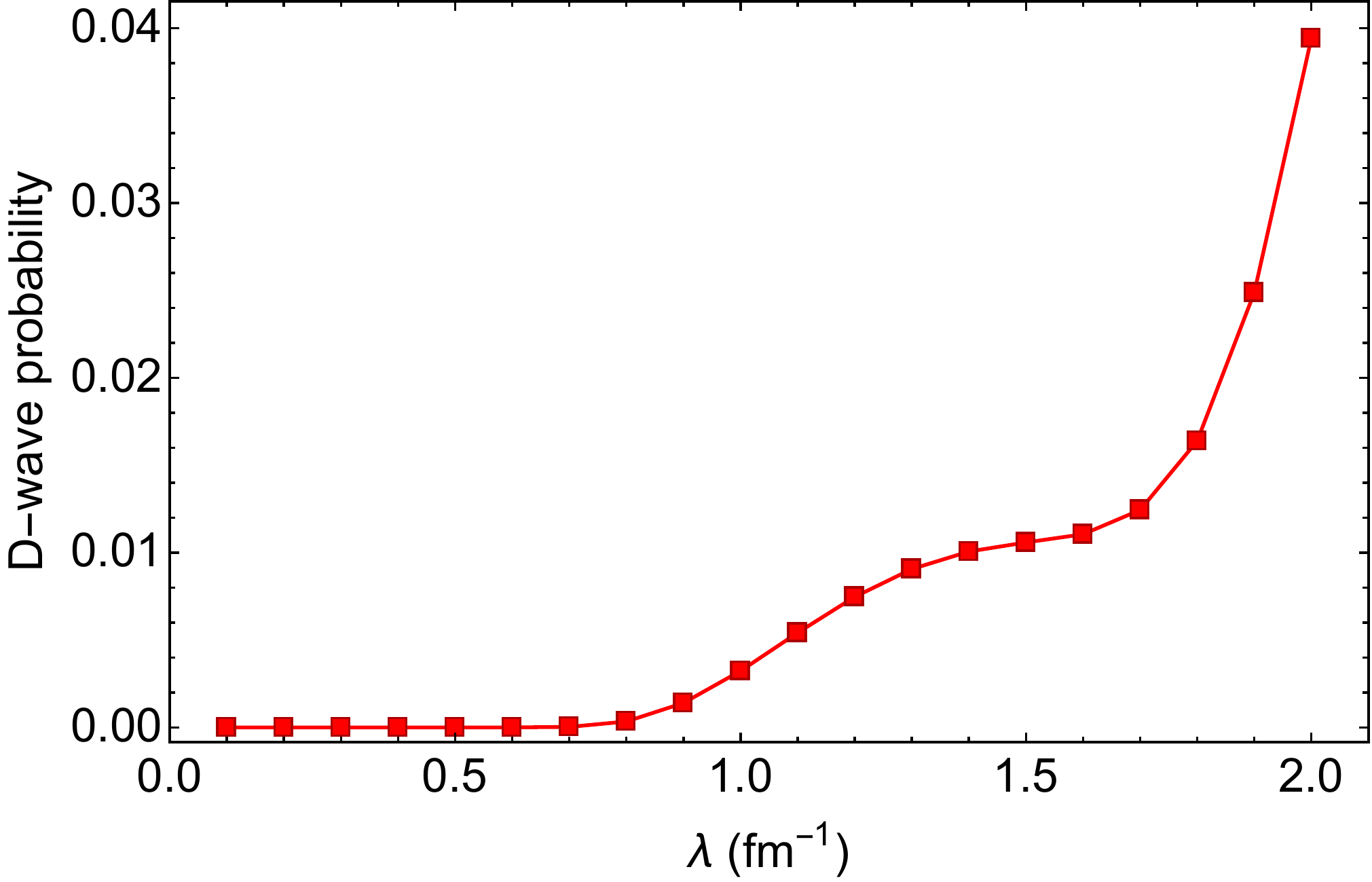}
\end{center}
\caption{Running of deuteron state probabilities with the SRG cutoff: S-wave (left) and D-wave (right).}
\label{fig7}
\end{figure*}

In Fig. \ref{fig7} we show the probability of finding the deuteron in the $^3S_1$ and $^3D_1$ states as a function o the SRG cutoff.
At $\lambda = 2.0 ~{\rm fm}^{-1}$ the probabilities are about 96\% and 4\% respectively as expected from an interaction that describes
two-nucleon observables with high accuracy. Once $\lambda$ goes down, the S-wave probability increases and the D-wave probabilities
decreases until $\lambda = 0.9 ~{\rm fm}^{-1}$ is reached. Below this point, the D-wave is suppressed and a pure S-wave deuteron emerges.

\begin{figure*}[h]
\begin{center}
\includegraphics[scale=0.4]{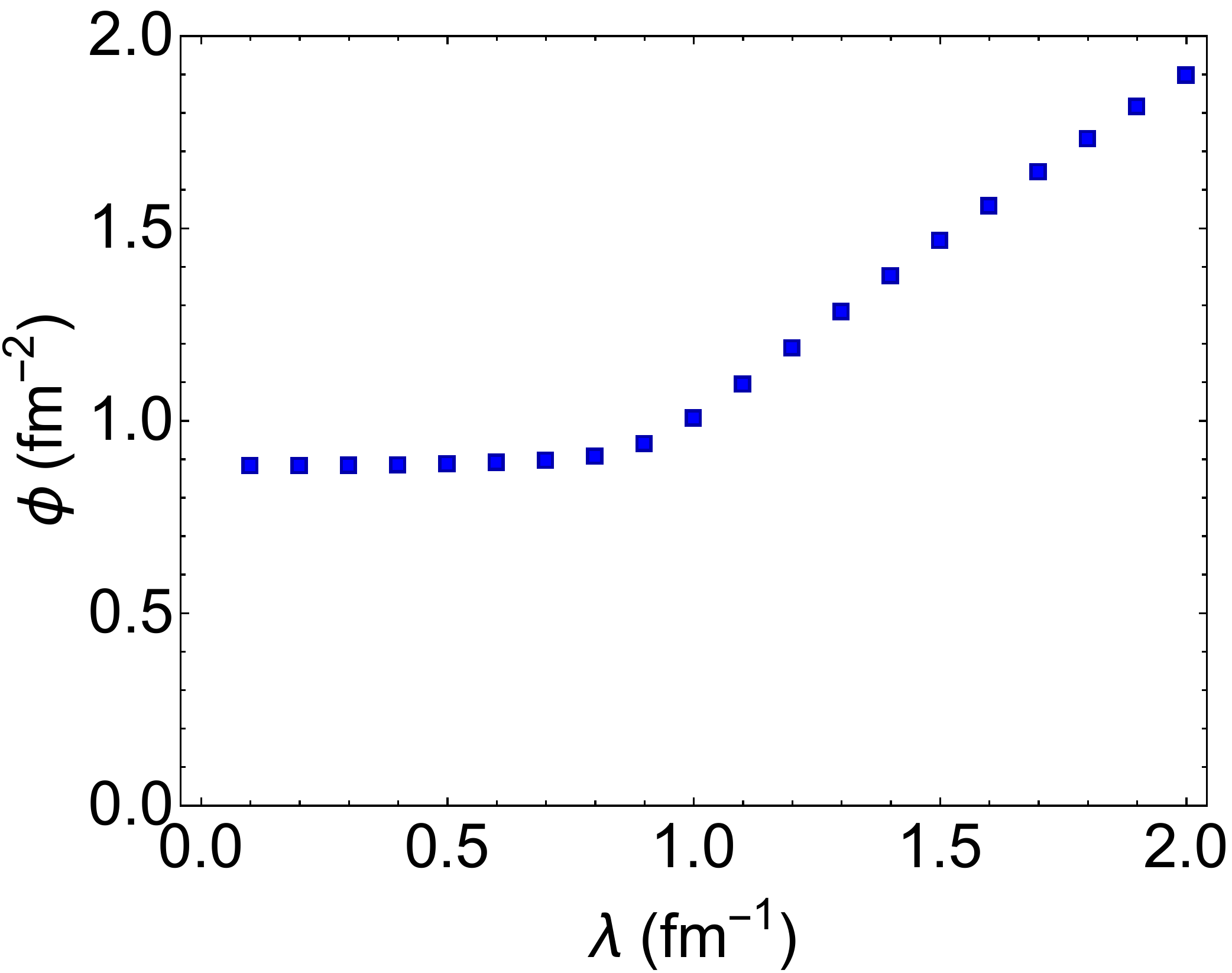} \\ 
\includegraphics[scale=0.4]{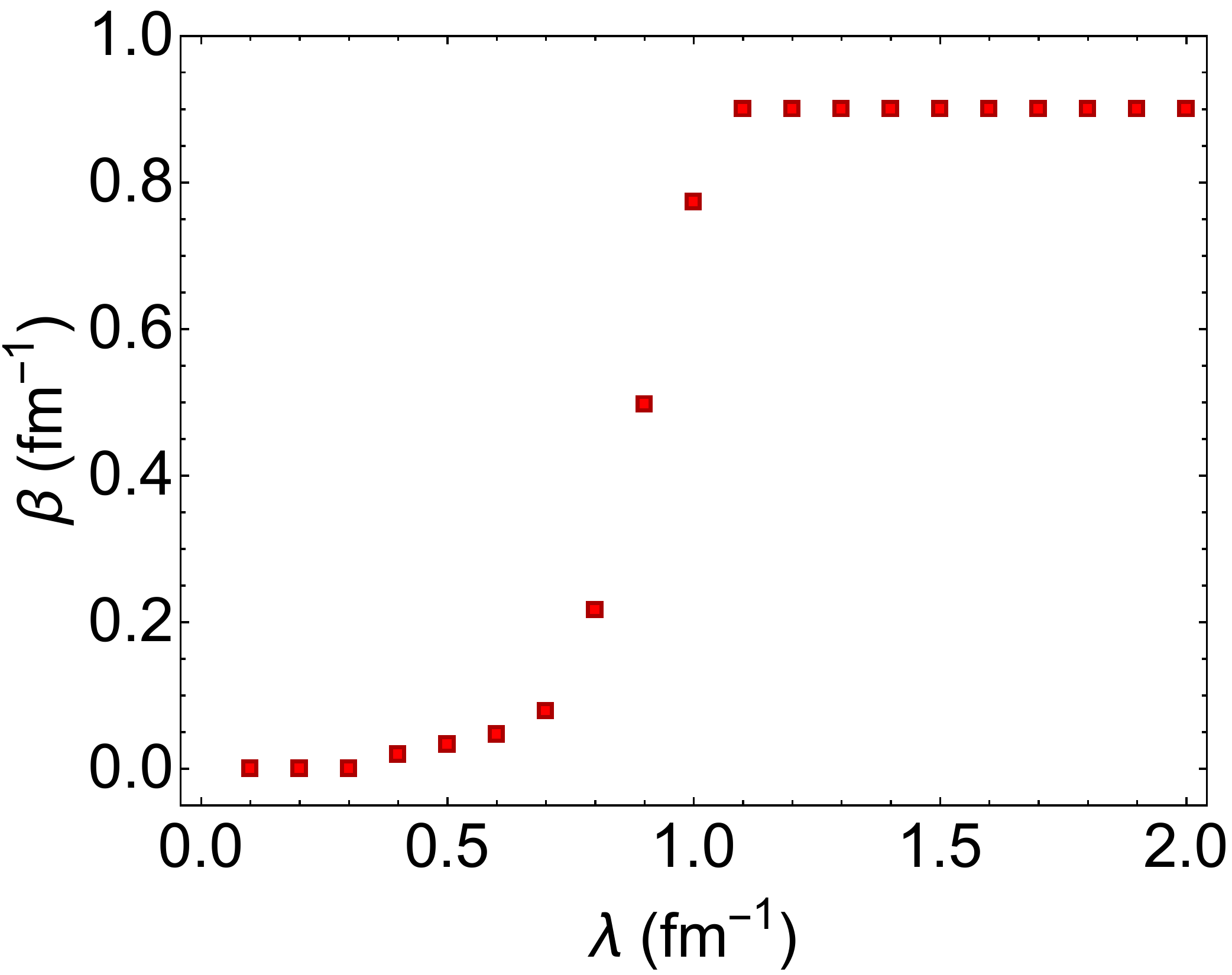} \\ 
\includegraphics[scale=0.4]{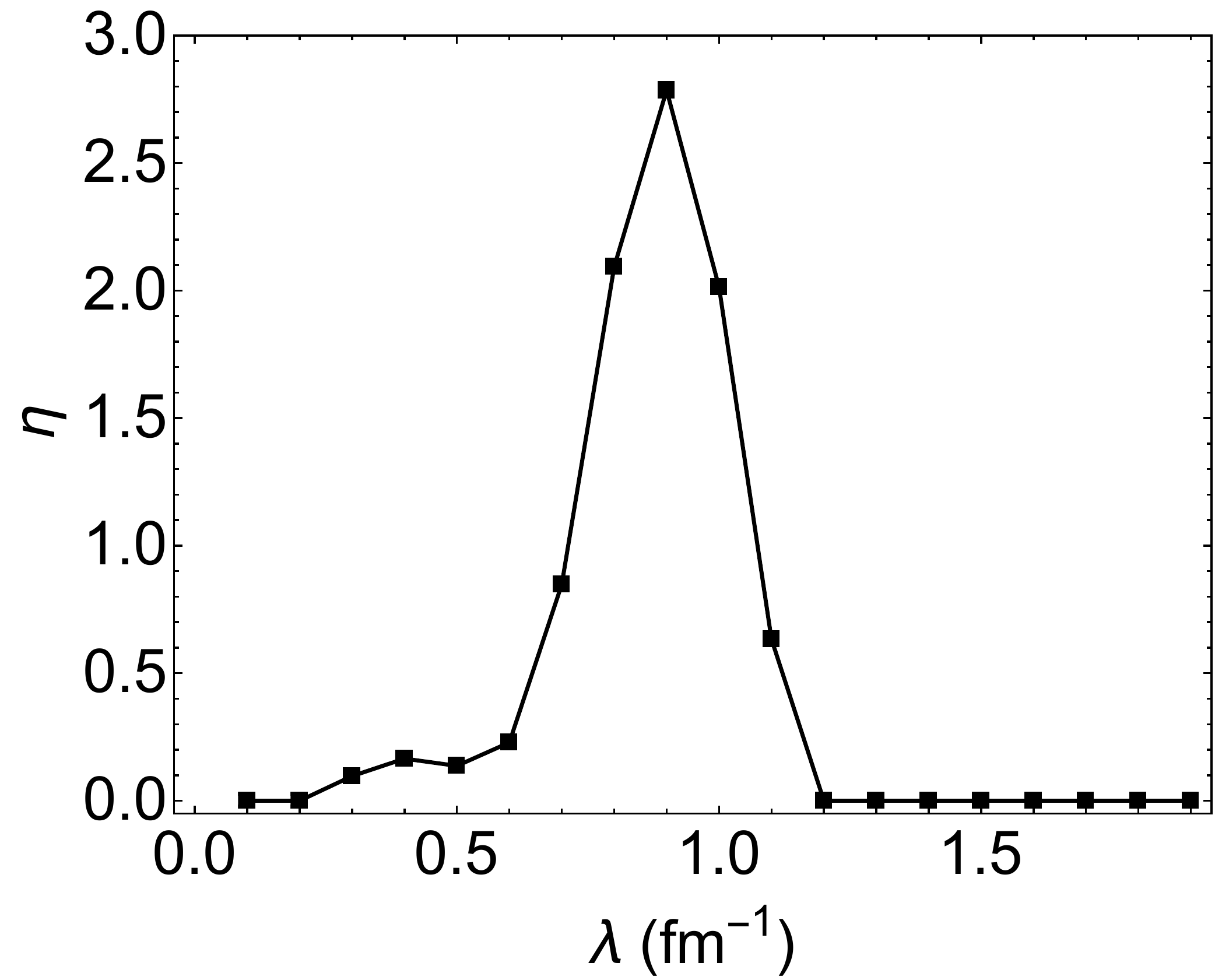}
\end{center}
\caption{The Frobenius norm $\phi$ of the N3LO potential matrix in the $^1S_0$ channel (left panel), 
the order parameter $\beta = \partial\phi / \partial\lambda$ (middle panel) and the dimensionless 
similarity susceptibility $\eta = \partial\beta / \partial\lambda$ (right panel) as a function of the SRG cutoff $\lambda$ 
for $N = 30$ grid points.}
\label{fig8}
\end{figure*}

The binding energies as well as the generalized Tjon lines present a turn over in their behavior as the two-nucleon interaction
approaches the on-shell limit. We observe that this is actually related to the off-shell / on-shell phase transition that occurs during the SRG flow. 
The running of the Frobenius norm $\phi$ and its derivatives with the SRG cutoff are shown in Fig. \ref{fig8} where we can observe 
a crossover behavior of the order parameter $\beta$ (center panel) as the norm stabilises when reaching the fixed point (left panel).
The peak in the similarity susceptibility $\eta$ (right panel) indicates that the off-shell / on-shell transition for the N3LO chiral potential 
occurs at $\lambda_c = 0.9~{\rm fm}^{-1}$ when the maximum momentum is $\Lambda = 4~{\rm fm}^{-1}$ and the number of points 
is $N=30$. Increasing the number of points makes the peak to move towards smaller SRG cutoffs and the continuum limit 
implies in $\lambda_c \to 0 $ \cite{Timoteo:2016vlp}. 

\section{Final remarks}

We explored the binding energies of light nuclei and the interplay between two-body and three-body contributions 
during the SRG flow of a chiral nucleon-nucleon interaction. By computing the two-body contribution and determining 
its running with the SRG cutoff, we were able to infer the running of the three-body contribution. For the $^3H$ the 
only assumption is that its binding energy comes from two-body plus three-body contributions and for the $^4He$ an extra
assumption that four-body contribution is negligible is required. 

Our calculations show that the on-shell transition during the SRG flow is what drives the change in the binding energies
and results in the back-bending of the generalized Tjon lines. The critical scale $\lambda_c = 0.9~{\rm fm}^{-1}$ seems
to separate two regimes where the effective interaction is predominantly off-shell ($\lambda > \lambda_c$) or on-shell 
($\lambda < \lambda_c$). Surprisingly, the same scale provided the best value for the Bertsch parameter starting from
a separable potential \cite{Arriola:2017xyz}.

While, the SRG scale has been traditionally regarded as an auxiliary mathematical device to soften nuclear interactions, 
our results point to a more physical understanding of the SRG renormalization scale which might hopefully be exploited 
in the future in large scale calculations, namely that it is possible to describe nuclear binding with almost on-shell interactions 
and moderate three-body forces.

Our aim for the moment is to achieve some conceptual understanding of the physics underlying the SRG flow, which has been 
often used to soften nuclear interactions. We of course would like to provide in the end simple prescription which might profitably 
be used in large scale calculations. As we mention, being in practice close to the on-shell situation is not only physically appealing 
but could be of numerical use, see e.g. Ref. \cite{Arriola:2015hra} for a practical implementation. We don’t expect to improve 
quantitatively over {\it ab initio} calculations which are currently the state-of-the-art, but we believe simple approaches are important 
to study some aspects of the SRG flow, in particular the infrared limit ($\lambda\to0$). And for qualitative purposes high numerical accuracy 
is not required. So, we view computationally expensive {\it ab initio} calculations and simple approaches as being complementary.

\section*{Acknowledgements}

E.R.A. would like to thank finantial support by the Spanish MINECO and European FEDER funds (grant FIS2017-85053-C2-1-P) 
and Junta de Andalucia (grant FQM-225). S.S. and V.S.T. would like to thank FAPESP (grants 2017/13282-5 and 2019/10889-1) 
for financial support. V.S.T. also thanks CNPQ (grant 306615/2018-5) and FAEPEX (grant 3258/19).

\clearpage

\end{document}